\documentclass[twocolumn,showpacs,preprintnumbers,amsmath,amssymb]{revtex4}

\usepackage{graphicx}
\usepackage{dcolumn}
\usepackage{bm}
\usepackage{psfrag}

\usepackage{color}

\begin{document}

\preprint{Submitted}

\title{Nonlinear effects for coda-type elastic waves in stressed granular media}

\author{V. Tournat}
\affiliation{LAUM, CNRS, Universit\'e du Maine, Av. O. Messiaen, 72085 Le Mans, France}
\email{vincent.tournat@univ-lemans.fr}
\author{V.~E. Gusev}%
\affiliation{Laboratoire de Physique de l'\'Etat Condens\'e, UMR-CNRS 6087, Universit\'e du Maine, Av. O. Messiaen, 72085, Le Mans, France}

\date{\today}

\begin{abstract}
Experimental results and their interpretations are presented on the
nonlinear acoustic effects of multiple scattered elastic waves in unconsolidated
granular media. Short wave packets with a central frequency higher
than the so-called cut-off frequency of the medium are emitted at one
side of the statically stressed slab of glass beads and received at
the other side after multiple scattering and nonlinear
interactions. Typical signals are strongly distorted compared to their
initially radiated shape both due to nonlinearity and scattering. It
is shown that acoustic waves with a deformation amplitude much lower
than the mean static deformation of the contacts in the medium can
modify the elastic properties of the medium, especially for the weak
contact skeleton part. This addresses the problem of reproducibility
of granular structures during and after acoustic excitation, which is
necessary to understand in the non destructive testing of the elastic
properties of granular media by acoustic methods. Coda signal
analysis is shown to be a powerful time-resolved tool to monitor slight
modifications in the elastic response of an unconsolidated granular
structure.
\end{abstract}

\pacs{Valid PACS appear here}

\maketitle

\section{Introduction}

Unconsolidated granular materials are known to exhibit a high complexity in their macroscopic behaviors. This gives rise to numerous fundamental and applied processes that are currently intensively studied, such as avalanches, dune formation, compaction etc \cite{jae92,cat98,deg99,ric05}. Most of the studied physical processes concerning the unconsolidated granular matter are obviously related to the possibility for the grains to move relative to each other, from microscopic (much smaller than a grain diameter) to macroscopic (much larger than a grain diameter) rearrangements. While large movements of grains have been observed and modelled for the processes where the granular material behavior may be fluid-like (even complex and not classical), like avalanches, microscopic relative movements are less intuitive and less studied. They are however the only possible relative movements in solid-like granular media where a static stress is applied and the grains are confined in a given volume. 

In this context, the acoustic waves, known to be highly sensitive to
the contact stresses \cite{liu92,liu93,tou04a,jia04}, can be useful
for the monitoring of these small relative movements \cite{zai05},
impossible to detect by other existing methods. They could even be
useful for the generation of these small movements. Among the several
existing methods for the grain movements or contact stresses
monitoring (carbon paper experiment, photo-elasticity, image
processing from CCD camera, X-ray tomography or photo-imaging in
iso-index configuration \cite{mue98,cat98,bla01,ric05}), the acoustic
methods are, to the best of our knowledge, the only ones sensitive to
the contact stresses inside the volume of a 3-D non transparent
medium. One of the main advantage of acoustic methods is also the
temporal resolution of the measurement, which could decrease down to
the wave period (one millisecond at one kHz), and is particularly
interesting in transient processes like avalanches for
instance. However, there is currently a lack of understanding of the
acoustic transport properties through granular media, which makes these methods mostly qualitative.

Several previous works \cite{jia99,tou04a,tou03,zai05} have shown the following qualitative features for the acoustic wave propagation through stressed and non cohesive granular media. The propagation of a low amplitude (with a strain much lower than the average static strain of the contacts) and low frequency (with a wavelength much larger than the grain diameter) wave can be linear and is mostly governed by the averaged properties of the medium (average density, average static stress, coordination number) \cite{God90,Mak04}. The propagation of a high amplitude (with a strain lower than the average static strain of the contacts but which can be comparable) and low frequency wave is nonlinear and governed partly by the so-called weak contacts (with a static strain lower than the average one) \cite{tou04a,tou04d,tou04b}. This is a consequence of the fact that individual stress-strain relationships for each contact have higher values of nonlinear parameters when the pre-stress is smaller. For acoustic waves with wavelengths comparable to the bead diameter, however, these simple features are not straightforward, and the transport properties determination remains a fundamental problem. How to describe the wave scattering in a network of beads with geometrical and contact disorder? What is the role of the so-called force chains in the transport properties of short wavelength acoustic waves? Can the weakest contacts make the multiple scattering process nonlinear?

A linear diffusion approach has been recently applied to the analysis of coda-type signals in confined granular media \cite{jia04}. This is one of the first attempt to understand the transport properties of short wavelength elastic waves in such media at the laboratory scale and to extract parameters such as a characteristic time of absorption $\tau_a$ and a diffusivity $D$. This approach is based on previous works in elastic wave diffusion carried out for slightly different media (solid rods in water \cite{tou00}, chaotic cavities \cite{wea82}, or glass bead in water \cite{sch97} for instance) and on a different scale \cite{aki75,hen01}. 

In this article, we report some experimental results of nonlinear
acoustic wave propagation, with initially radiated wavelengths close
to the bead diameter. In some well-chosen experimental configurations,
it is possible to observe coda-type signals, together with a nonlinear
signal of low frequency, which is identified here as a
self-demodulated signal. For some widely encountered configurations,
linear and nonlinear wave processes of multiple scattering,
self-demodulation and self-action are observed \cite{tou05c,tou07b}. The last nonlinear process of self-action of the multiple scattered field is observed even for moderate excitation amplitudes (deformations much lower than the average static strain of the contacts). The application of some of the presented effects to the characterization of granular media is finally discussed. 

\section{\label{sec:exp}Experiments}

As a preamble, carrying quantitative experiments on acoustic wave propagation in unconsolidated granular media is a hard task for several reasons that are described in the following. 

First, in order to excite the modes of propagation inside the solid
frame, i.e. through the elastic beads and their contacts, and not
through the fluid saturating phase, it is necessary to put the
piezo-transducer in direct contact with the medium itself (or in
contact with a solid which is itself in direct contact with the
medium). Other types of transducers for the excitation of acoustic
waves, like laser generation of ultrasound for instance, have not been
applied to our knowledge, certainly because of the high efficiency
needed in these strongly attenuating media. The application of a
mechanical stress with a contact transducer has several consequences
on the possible experiments. What can be done usually in liquids for
acoustic metrology, i.e. changing the emitter-receiver distance in
order to measure the wave velocity for instance, is extremely
difficult in unconsolidated granular media, because the contact
network is modified by any variation of external conditions.

The second difficulty in carrying quantitative experiments is the
experimentally observed strong influence of the applied static stress
on the transmission coefficient especially for short wavelength
acoustic waves. This result is discussed in the following. The
consequence is the impossibility to perform quantitative reproducible
and precise experiments if ones modify the applied static stress on
the sample, even if it is measured to be equal after a back and forth
modification. Moreover, if one would be able to apply exactly the same
stress, the memory effects due for instance to the hysteresis in the
quasi-static stress-strain relationship, take place
\cite{hol81,guy95,jos00}. In an attempt to control carefully the
granular preparation, the acoustics of compaction has been recently
studied \cite{ins08}. However, it allows to perform experiments only
for weakly stressed granular packings (typically from $75\ Pa$ to $3\ kPa$), far from the stress necessary for the observation of the signals described here.

The third difficulty, is related to the fragility of such media
\cite{cat98,jae92}. Due to the inhomogeneity of the individual contact
loads \cite{mue98,bla01,eri02,Mak04}, some of the contacts are weakly
loaded, which make them able to clap, slip or slide under the action
of a small amplitude strain wave. In contrast, the so-called strong
contacts, are supposed to take part in the force chains which
constitute the elastic skeleton of the bead arrangement. Previous
works show by simple analytical formulae or experimentally that the
weaker the contact the higher is its acoustic nonlinearity
\cite{tou04a}. It is thus expected that the major role in the linear
acoustic properties is played by the strong contacts and the force
chains while the nonlinear acoustic properties are more controlled by
the weak contacts. There exist few experimental configurations (in
particular when short wavelengths are used) when these roles cannot be
qualitatively separated like that, and either weak contacts influence
importantly the linear properties \cite{cos08} or strong contacts are involved in
the observed nonlinear effects.

\subsection{{\label{subsec:expset}Experimental setup}}

\begin{figure}
\begin{center}
\includegraphics[width=8.5cm]{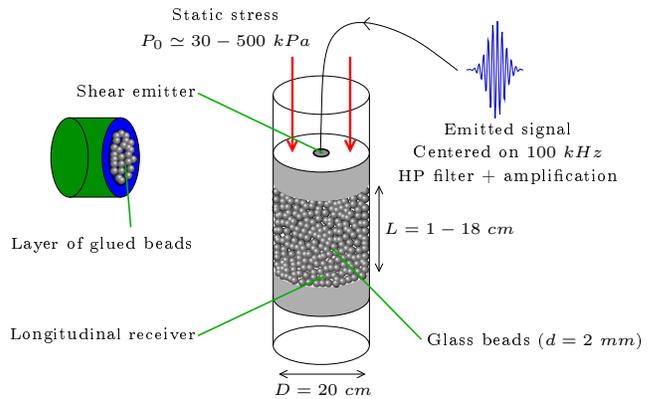}
\caption{Schematics of the experimental configuration.}
\label{fig1}
\end{center}
\end{figure}

In Fig.~\ref{fig1}, a schematic representation of the experimental
setup is shown. A wave packet made of a sine wave at $100 \ kHz$
modulated in amplitude by a Gaussian function is high-pass filtered
over $40 \ kHz$ (in order to avoid any direct low-frequency
excitation), amplified and launched in the medium with a shear
piezo-transducer. Due to multiple scattering,
mode conversions (including dilatancy), the acoustic energy is rapidly distributed
between longitudinal and shear vibrations in the medium. The use of a
shear emitter has the advantage of avoiding the direct excitation of a
pressure wave in the air saturating the beads. In the following results, two types of Gaussian
function widths for the initially launched wave packets have been
used, and two types of surface treatment for the shear transducer (the
first type is the raw transducer membrane with asperities estimated to
be of characteristic scale $\sim 50 \ \mu m$ and the second type is a
transducer with a glued single layer of $2 \ mm$ diameter glass beads,
the same beads as in the medium itself). The container diameter
is $D=20 \ cm$ and the propagation length ranges between $1\ cm$ and
$18\ cm$. The applied static uniaxial stress is measured with a force
sensor at one end of the container and ranges from $30\ kPa$ to $500\
kPa$. A longitudinal piezo-electric receiver is placed at the bottom of the
container. This transducer has been chosen because of its high sensitivity
over a wide-frequency band, allowing for a large range in the acoustic
amplitude measurements. In all the experiments presented in the
following, the estimated average contact strain $\varepsilon_0$
(of the typical range $3 - 7 \times 10^{-4}$ for a $200\ kPa$ applied stress) is always greater than
the maximum acoustic strain $\varepsilon_{max} \simeq 10^{-5}$.

When using this setup, it is possible to modify the emitted acoustic frequency, the acoustic amplitude, the applied static stress, the propagation distance. But importantly, it is better for a quantitative insight in the wave propagation phenomena, to modify parameters of the setup that change the smallest number of medium (or wave) parameters. For instance, changing the frequency or the static stress modifies the wave scattering, absorption, dispersion and nonlinearity. Changing the distance, changes irreversibly and in an uncontrolled way the granular medium state, as it is necessary to unload completely the medium and add or remove some granular matter. The convenient parameter to modify is the wave excitation amplitude, which can be accurately measured, and is reproducible. The medium is a priori not irreversibly modified.

In the following several experimental results are presented and their implications on the granular material elasticity is discussed. The parameters of the experiment may be different, they are consequently recalled for each group of results when necessary.

\subsection{Multiple scattering and nonlinearity}

\begin{figure}
\includegraphics[width=8.5cm]{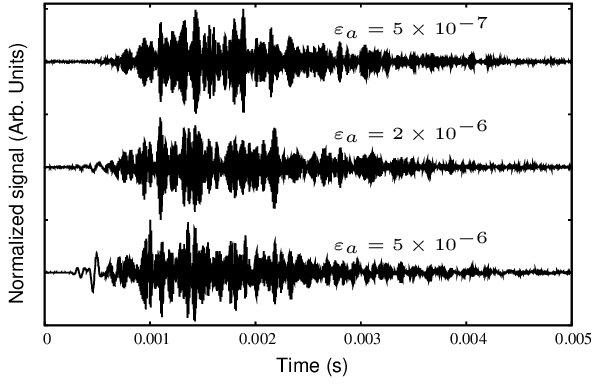}
\caption{Experimental temporal signals that are typically received for
  different excitation amplitudes. The applied static stress is $\sim
  300\ kPa$ and the emitter - receiver distance is $D = 10.5\ cm$.}
\label{fig2}
\end{figure}

The signals presented in Fig.~\ref{fig2} are typical experimental signals that can be observed when the initial acoustic wavelength is of the order or less than the bead size and for a static stress larger than $\sim 100\ kPa$). For an estimation of the wavelength, one can use the elastic parameters of glass and the stress-strain relationship of an average loaded contact in an effective medium theory. However, it is important to keep in mind that this definition neglects wave velocity dispersion which can be important when the spatial scale of the beads coincides with the wavelength. As an estimation, the wave velocity in a disordered three-dimensional packing of $2\ mm$ glass beads with an applied static stress of $200\ kPa$ is $\sim 300 \pm 40\ m/s$, which gives a wavelength of $\sim 3 \pm 0.4 \ mm$ at $100 \ kHz$.
This means that for a wave packet centered on $100 \ kHz$, submitted
to a $200\ kPa$ static stress, the wavelength is of the order of the
bead diameter, and consequently, strong scattering occurs. The typical
experimental signals received in this case are strongly distorted due
to multiple scattering, lasting from $10$ to $100$ times longer than
the emitted signal (see Fig.~\ref{fig2}). This feature is widely
observed in seismics \cite{aki75} or in multiple scattering
experiments, and has been observed in roughly the same conditions in
glass bead assemblage \cite{jia99}. However, we found another feature
in this case, not mentioned previously in the literature: for
different excitation amplitudes, the normalized signals in
Fig.~\ref{fig2} are different. This relative amplitude and shape
dependence is the signature of a nonlinear process. One obvious
difference between the traces is the emergence, for an increasing
excitation amplitude, of a signal at earlier times, relatively lower
in frequency than the $100 \ kHz$  coda-type signal. We verified in
several other experiments \cite{tou04a,tou03} that this LF signal is
the self-demodulated contribution associated with the rectification
(demodulation) of the initial HF wave packet. Its nonlinear nature is
evidenced by its nonlinear amplitude dependence on the excitation
amplitude. This is an important difference compared to results
\cite{jia99,jia04}, where the very same types of signals were observed
(for instance for $\varepsilon_a = 5 \times 10^{-6}$ in Fig.~\ref{fig2}). In
\cite{jia99}, the LF contribution is identified as the linear coherent
part of the propagated initial pulse, and the HF contribution is
interpreted as the multiply scattered signal part. A possible
explanation for such a different observation (mainly in the amplitude
dynamics behavior and physical nature of the LF signal) is the
difference in the spectral width of the emitted signal spectra. In
Ref.~\cite{jia99}, the signal spectrum width is relatively wide,
which may allow for the direct radiation of the linear coherent LF
component in the medium. Then, the observation may be interpreted as a strong frequency dependent transmission in the medium (due to absorption and scattering), which is able to modify importantly the transport character of different frequencies (propagative as in an effective medium, multiply scattered, diffusive...). 

\begin{figure}
\begin{center}
\includegraphics[width=8.5cm]{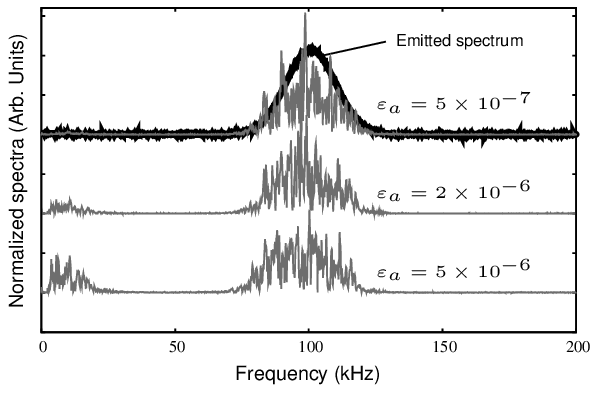}
\caption{Experimental spectra corresponding to temporal signals of Fig.~\ref{fig2}. The spectrum of the emitted signal with a $22 \%$ FWHM (full width at half maximum) is superimposed on the top curve.}
\label{fig3}
\end{center}
\end{figure}
In the present case however, when the emitted spectrum is sharper, the
direct radiation of the LF signal is negligible (and in any case is avoided by a high-pass filter in the setup). This LF signal is generated in the medium itself, by nonlinear processes of frequency mixing \cite{tou04a}, widely described in the literature \cite{wes63,nov87,zve99,zai99,zai99b}. This can be seen in Fig.~\ref{fig3}, where the relative amplitudes of the LF and the HF contributions are modified by increasing the excitation amplitude. 

\begin{figure}
\begin{center}
\includegraphics[width=8.5cm]{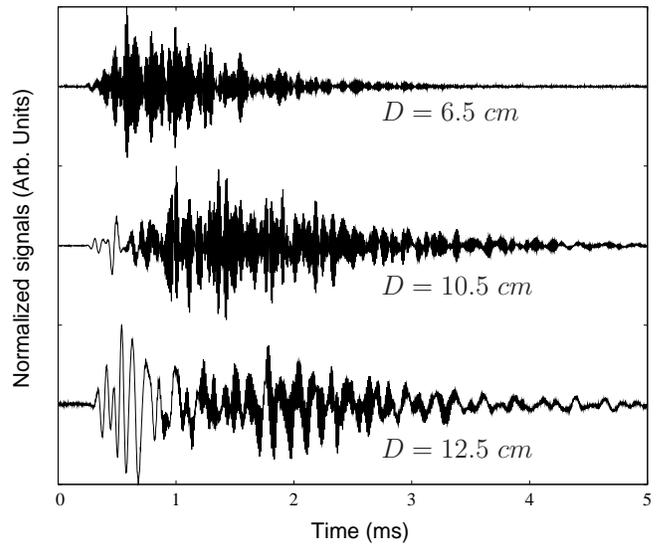}
\caption{Signals received after propagation through different
  thicknesses of granular medium, for identical static pressure of 300
  kPa and excitation amplitude $\varepsilon_a=5 \times 10^{-6}$.}
\label{fig5}
\end{center}
\end{figure}

In Fig.~\ref{fig5}, the received temporal signals are presented for
three source-receiver distances, with the same applied static stress
$300\ kPa$ (measured with a precision of $\sim 3\ \%$ which is not
sufficient to perform precise quantitative measurements). An important
qualitative feature observed on these normalized (by their maximum)
traces is the relative increase of the LF contribution compared to the
HF one with distance. This feature is usually observed for the
nonlinear self-demodulation process \cite{nov87}, where over a
sufficient distance, only the LF contribution is detectable, due to
the difference in attenuation of the HF and the LF waves. In this
experiment, the LF signal absolute energy is measured to be roughly
constant with distance, while the received HF signal energy is
decreasing drastically. The former is a result of the competition between
nonlinear effects, which pump low frequencies through frequency-down-conversion processes, and linear attenuation.

\begin{figure}
\begin{center}
\includegraphics[width=8.5cm]{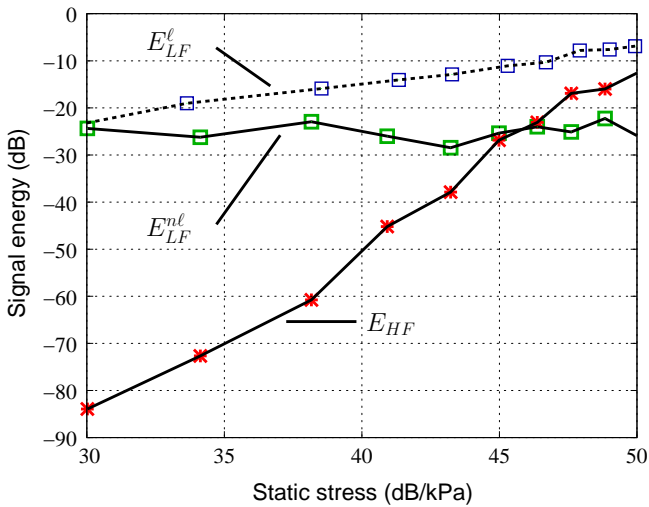}
\caption{Signal energy dynamics as a function of the applied static pressure. The HF ($E_{HF}$) and LF ($E_{LF}^{n\ell}$) contributions are separated by filtering. The energy of a reference LF ($E_{LF}^{\ell}$) linear signal is also plotted for comparison.
}
\label{fig6}
\end{center}
\end{figure}

In Fig.~\ref{fig6}, the total energy of different signals is plotted
as a function of the applied static stress for a propagation distance
$D=12.5\ cm$ and an excitation amplitude $\varepsilon_a \simeq
\varepsilon_{max}/50$, where $\varepsilon_{max} \simeq 10^{-5}$. Three
signal energies are defined here: the energy of the HF contribution
$E_{HF}$, the energy of the LF contribution $E_{LF}^{n\ell}$ and the
energy of a directly radiated LF signal $E_{LF}^{\ell}$ (in this case,
a LF pulse with the same frequency content as the nonlinearly demodulated pulse is directly
excited by the emitter and received after linear propagation). First,
comparing the received energies $E_{HF}$ and $E_{LF}^{\ell}$ evolution
as a function of the applied static stress, an important slope
difference is clearly visible: $E_{LF}^{\ell}$ dependence on stress is
less than to a power 1 and $E_{HF}$ dependence is close to a power 4
of the stress. In order to explain this difference in the evolution of
the transmitted energy as a function of the applied static stress, one
should recall that two dominant processes of sound attenuation may
play a role: the scattering and the absorption due to linear processes. Considering that the scattering of the HF $\sim 100\ kHz$, is much more important than for the LF $\sim 10\ kHz$, the observed difference may be attributed to the strong stress dependent scattering. This is consistent with the simple idea that by increasing the static stress, contacts are created which adds new paths (or strengthen existing paths) of acoustic energy transmission. Moreover, waves following force chains with a given stress may change from an evanescent character to a propagative one with increasing stress, because the cut-off frequency (sometimes called the Einstein frequency) is increased \cite{gil03,cos08,tou04c}. It is then possible to ask if this scattering which strongly depends on static stress depends also on the acoustic amplitude of excitation. Is the acoustic wave able to switch dynamically propagation paths or force chains? The aim of the next section is to analyze this opportunity and its possible manifestations. 

The competition between nonlinear effects and attenuation is also visible in Fig.~\ref{fig6} when comparing the dynamics of $E_{LF}^{\ell}$ and $E_{LF}^{n\ell}$. For signals having the same low frequency propagating through the same medium but generated either directly by a transducer or by nonlinear effects inside the medium itself, the energy dynamics as a function of the applied static stress is different. It can be assumed that the linear attenuation varies in the same way for both signals while the nonlinearity of the medium and the HF attenuation both playing a role in the generation of the LF signal \cite{nov87} may be strongly affected. These modifications only affect the nonlinearly generated signal. The $E_{LF}^{n\ell}$ dynamics as a function of the applied static stress is almost constant, while the $E_{LF}^{\ell}$ is increasing. It means that the efficiency of the nonlinear interactions leading to the self-demodulated signal is decreasing with increasing static stress. Two main processes can lead to a diminishing of the nonlinear interaction efficiency: the HF attenuation increase (the amplitude of the nonlinear sources diminishes) and the intrinsic nonlinearity of the medium decrease. As the HF attenuation (probed by the dynamics of $E_{HF}$) is strongly diminishing with increasing static stress, it means that the nonlinearity of the medium is decreasing with increasing static stress. This is consistent with estimates based on the Hertzian contact nonlinearity and experimental observations \cite{tou04a,tou03}.

Now, we analyze the temporal features of the HF signals, namely the
codas. Coda features can be analyzed by performing an envelope
detection (for instance by integrating the square field over a sliding
window). This may cause problems for quantitative analysis because one
has a degree of freedom for the size of the sliding window. In
principle, the diffusion approximation is shown to be valid for the
configurational average intensity $\langle I \rangle$ of the
incoherent field under some conditions \cite{der01b}. The incoherent
field can be defined as the contribution which cancels out with
averaging over different realizations of the disorder in the medium
\cite{der01,der01b}. The average intensity $\langle I \rangle$ is
usually accessed experimentally by the use of an intensity sensor,
which is large compared to the wavelength (a photo-multiplier in
optics for instance). For elastic waves however, the piezo-electric
transducers are sensitive to the field itself. It is thus necessary,
in order to record $\langle I \rangle$, to record wave field signals,
then to calculate the corresponding intensities, and finally to average the different obtained intensities over different configurations of the disorder. However, when the receiving transducer has a sensitive surface with dimensions larger than the wavelength, a spatial averaging is performed on the field. In this operation, the incoherent contribution is diminished compared to the coherent one. Consequently, even if most of the energy is incoherent at the receiver location, there could be some situations where the detected signal is composed of coherent and incoherent parts of comparable importance.

In unconsolidated granular media, the configurational averaging can be
performed by several means, which may not be equivalent. One could use
arrays of transducers for the excitation and detection of the multiple
scattered waves, each transducer being small compared to the
wavelength. Then without modifying the medium, each emitter - receiver
couple, for the same distance through the medium could provide a
signal to be treated. Otherwise, with one emitter and one receiver,
one could modify the structure of the medium (geometrical and contact
force disorder) by keeping constant as much as possible the
macroscopic external conditions \cite{jia04}. These two means can
allow in some limits for an estimation of the diffusivity and the
absorption characteristic time of the diffusion equation. The first
method necessitates small enough transducers to avoid averaging of the
field, which causes problems of coupling with the medium when the
wavelength is of the order of the bead diameter and the transducer
size. Some of the transducers may not be in efficient contact with
beads. The second method, due to the memory effects \cite{jos00,mou01}
or the drastic influence of the external parameters on the elastic
properties (see Fig.~\ref{fig6} for instance), does not guarantee to
be in the same experimental conditions from one realization of the
disorder to another. Small perturbations or local modifications of the
contacts forces in the medium are consequently difficult to probe
using averaged quantities, obviously more adequate for an estimation of average properties of the medium.

\subsection{Analysis of the amplitude dependent codas}

In this section, we investigate fast dynamic modifications of the medium submitted to an acoustic wave, as well as slow dynamic modifications or permanent modifications of the medium. The experimental configuration is slightly different from the one in the previous section. Some glass beads have been glued to the shear emitting transducer surface, in order to couple better to the medium and to have a better signal to noise ratio.

\begin{figure}
\begin{center}
\includegraphics[width=8.5cm]{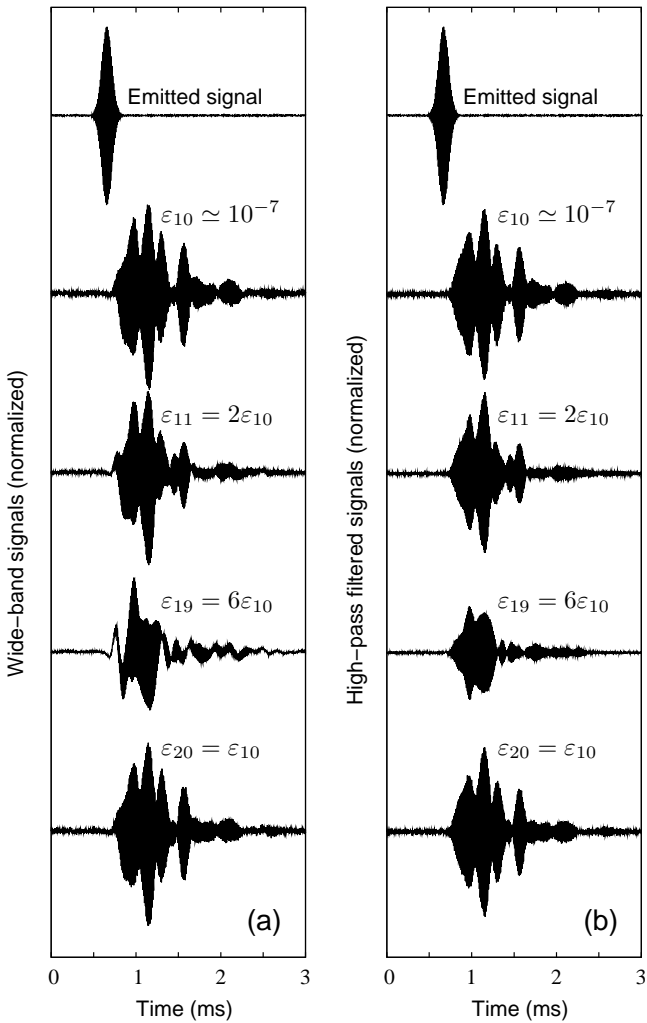}
\caption{Typical temporal signals (normalized by the excitation
  amplitude) that are received at a distance of $6\ cm$ from the
  emitter for different excitation amplitudes. The emitted signal is a
  Gaussian wave packet centered on $100\ kHz$ with a full width at
  half maximum (FWHM) of $6 \%$ in frequency. a) Row signals, b)
  High-pass filtered signals.}
\label{fig7}
\end{center}
\end{figure}

The analysis of the obtained spectra indicates that it is impossible
to detect neither the second nor the third harmonic component, certainly due to the strong observed attenuation above $100\ kHz$. However, the self-demodulated wave is quite easily generated and detected because of its lower attenuation. In the following, the HF coda contribution and this LF self-demodulated contribution are separated by post-processing filtering in order to analyze in details the different amplitude dependent processes.

\begin{figure}
\begin{center}
\includegraphics[width=8.5cm]{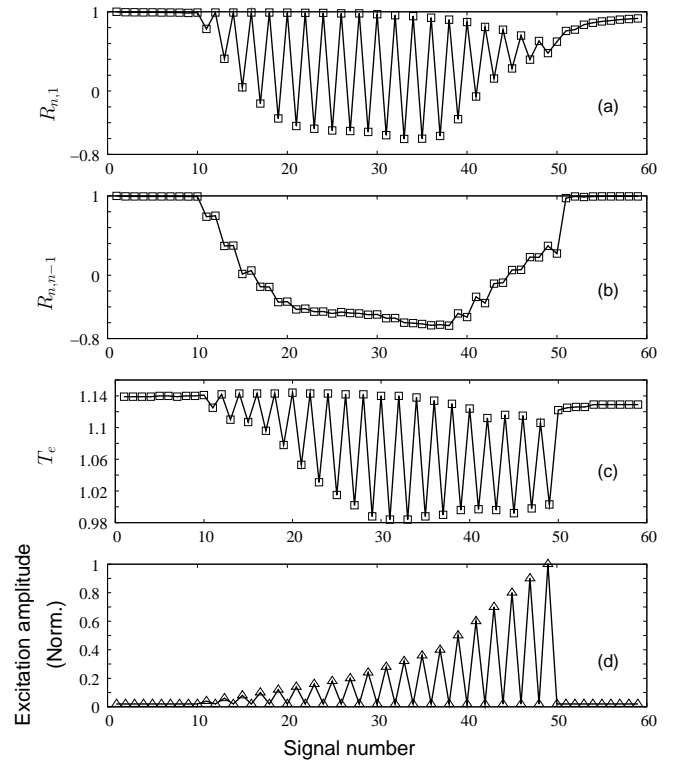}
\caption{(a) Resemblance parameter $R_{n,1}$ between the signal number $n$ and the signal number $1$ of the protocol (d). (b) Resemblance parameter $R_{n,n-1}$ between the signal number $n$ and the signal number $n-1$ of the protocol (d). (c) Arrival time $T_e$ of half the signal energy as a function of the signal number in the protocol (d). (d) Protocol for the successive excitation amplitudes of the experiment.}
\label{fig8}
\end{center}
\end{figure}

In Fig.~\ref{fig7}(a), typical signals that are received by a
sensitive wide-band longitudinal transducer (with a sensitive surface
of $4.5\ cm$ in diameter) are presented. The experimental protocol
elaborated in order to analyze the amplitude dependent effects of
acoustic wave transport is presented in Fig.~\ref{fig8}(d). It
contains 59 successive excitation amplitudes. The 10 first amplitudes,
as well as the last 10 amplitudes, are identical and relatively low
(the acoustic strain is $\varepsilon_a = \varepsilon_{1-10} \simeq
10^{-7}$). The odd numbered amplitudes 11-49 are gradually increasing
up to a maximum amplitude of $\varepsilon_a \simeq 10^{-5}$, while the
even amplitudes 10-50 are equal to the minimum probe amplitude
$\varepsilon_a \simeq 10^{-7}$. This allows to compare  the response
of the medium for different excitation amplitudes, but also, to
visualize the medium modifications after being excited by a strong
wave by using the weak probe wave. The experiment time for the
acquisition of one signal of this protocol is close to $30\ s$. In
Fig.~\ref{fig7}, signals are numbered according to their excitation
amplitude in the protocol, the first signal of the protocol being
denoted by $s_1$. The excitation signal is a sine wave modulated in
amplitude with a Gaussian function of $6\%$ full width at half maximum
(FWHM) in the frequency domain, which, at the central frequency
$f_0=100 \ kHz$ gives a $\sim 0.2 \ ms$ duration pulse at half the maximum. This duration is roughly the time necessary for the wave to travel from the emitter to the receiver for the $6\ cm$ distance.

In Fig.~\ref{fig7}(a), received signals over the full available
frequency band (from $1\ kHz$ to $300\ kHz$) are presented for several
excitation amplitudes of the protocole: $\varepsilon_{10}$,
$\varepsilon_{11}$, $\varepsilon_{19}$, $\varepsilon_{20}$. The signal
structure is quite complicated and shows several packets typically
4-5, having a shape close to the emitted signal. It is important to
note here that the time dependent intensity of these signals is much
different from a diffusive type enveloppe, even if the average
wavelength is of the order of the bead radius. Another important
feature is the emergence, at a sufficiently high excitation amplitude
of a relatively low frequency (LF) signal (see for instance signal
$s_{19}$ corresponding to the amplitude $\varepsilon_{19}$ in Fig.~\ref{fig7}(a)). Due to
its amplitude dependent character (this signal is not visible for
lower excitation amplitudes), this LF signal is nonlinear, and has
been identified in previous works as being a self-demodulated contribution \cite{tou04a,tou03}.

In the following, we analyze in details the high frequency part of the signals, and for quantitative comparison, the raw signals are high-pass filtered (Fig.~\ref{fig7}(b)). These coda-type signals have comparable time durations to those of Fig.~\ref{fig2}, but their envelope characteristic frequency is lower. This is a consequence of the narrower Gaussian spectrum of excitation. However, despite of the difference in coda shape, the characteristics of wave transport are similar. The most important characteristics of the signal evolution with excitation amplitude is the fact that their temporal shape is strongly modified, even for very moderate changes in the excitation amplitude. This can be seen by comparing the shapes $s_{10}$ and $s_{11}$ in Fig.~\ref{fig7}(b). Of course, by increasing further the excitation amplitude, the distortion is further increased (signal $s_{19}$). When the excitation amplitude is decreased down to the lowest amplitude of the protocole, i.e. to the probe amplitude $\varepsilon_{1}$, the shape modification of $s_{20}$ compared to $s_1$ is barely noticeable. In order to monitor quantitatively the modifications of the signal shapes due to amplitude dependent effects, we make use of the following parameter $R_{m,n}$ defined from the inter-correlation function $C_{n,m}(t)=\int_0^T s_n(t)s_m(t-\tau) d\tau$, where $T$ is the observation window duration, by
\begin{equation}
R_{n,m}=\frac{C_{n,m}(0)}{\sqrt{C_{n,n}(0) C_{m,m}(0)}} \ .
\end{equation}

The energy of signal $s_n$ (respectively $s_m$) is proportional to $C_{nn}(0)$ (respectively $C_{mm}(0)$). The normalization by $\sqrt{C_{nn}(0) C_{mm}(0)}$ makes consequently the parameter $R_{n,m}$ independent of the actual energies of $s_n$ and $s_m$, which is important in our case where signals with different amplitudes are compared. The parameter is only dependent on the relative shapes of the signals and can be seen as the level of resemblance of two signals, equal to 1 if the two signals are exactly the same and progressively diminishing when the signals differ.

In Fig.~\ref{fig8}(a), the parameter $R_{n,1}$ is plotted for each
signal $s_1$ to $s_{59}$ of the experimental protocol. It is clearly
visible that the parameter $R_{11,1}$ goes under a value of 0.9 while
the signals $s_2$ to $s_{10}$ exhibit a very accurate correlation with
$s_1$ ($R_{2-10,1} \simeq 1$). When the amplitude is increased
further, the parameter $R_{n,1}$ decreases drastically to 0 or less,
which shows that a strong field modification occured (the negative
values of the parameter $R_{m,n}$ do not have a particular meaning,
except when the value reaches $-1$, which means that the signal phases
are inverted). Concerning the even signals generated with a low
excitation amplitude, their resemblance parameter remains practically
equals to $1$ up to the 30th signal of the protocol. This means that
no modification of the granular medium elasticity is detectable by
this acoustic probing. In contrary, when higher acoustic amplitudes
are generated in the medium, the level of resemblance of the probe
wave with the first signals of the protocol begins to fall down
progressively, until the typical value of 0.6 after the maximum
excitation strain $\varepsilon_{49}$. The acoustical properties of the
medium have been modified by the strong wave action. A possible interpretation
of this experimental observation could be found in the temperature
effects on the linear and nonlinear hysteretic properties of the
medium \cite{gus05}. The wave action on the contacts increases the
average temperature and consequently modifies the linear and nonlinear
contact properties for some characteristic time which can be of the
order of several acoustic periods. Elastic memory effects at the
level of the contacts themselves (stick-slip motion for Hertz-Mindlin
contacts for instance \cite{min53}) or at the mesoscopic scale \cite{jos00} could also contribute. 

When, at the end of the protocol, only the probe amplitude is generated (signals $s_{50}$ to $s_{59}$), the parameter $R_{n,1}$ tends to 1 with time, the acoustic properties of the medium are progressively recovered in a slow healing process. The parameter $R_{n,n-1}$ in Fig.~\ref{fig8}(b) shows the level of resemblance of two successive signals. It is interesting to see that in this healing process at the end of the protocol, two successive signals are practically identical. The process of healing is slowly cumulative. This effect can be seen as a slow dynamic effect observed for elastic waves in other materials too, such as sandstones and cracked solids \cite{joh05,zai03}: the strong wave action modifies the acoustic properties of the medium, which are slowly recovered after some time, long compared to the wave period. The probing of this slow process is sometimes a powerful indication of damage (associated with the presence of internal contacts) in materials \cite{ben06}.

The last parameter plotted in Fig.~\ref{fig8}(c) is the characteristic
time $T_e$ corresponding to the arrival time of half the total energy
of the signal. It exhibits a quite clear amplitude dependence and has
the same qualitative behavior as the parameter $R_{n,1}$. Until the
30th amplitude of the protocol ($\varepsilon_{33}\simeq 0.3 \times 10^{-5}$), the characteristic time $T_e$ of the probe wave (even numbers of signals) remains unchanged, while the signals $s_{11}$ to $s_{31}$ exhibit a diminishing characteristic time $T_e$. It means that either the wave packet is propagating faster (nonlinear hardening of the medium), either the signal energy is preferentially attenuated at later times. Due to the strong nonlinear attenuation observed in the temporal signals of Fig.~\ref{fig7}(b), the second process seems more adequate to explain the $T_e$ dependence on amplitude. This is visible in Fig.~\ref{fig7}(b) when comparing signals $s_{10}$ and $s_{11}$ or $s_{10}$ and $s_{19}$. No shift in time is detectable while the last part of the signals are strongly attenuated relatively to their first part.

\begin{figure}
\begin{center}
\includegraphics[width=8.5cm]{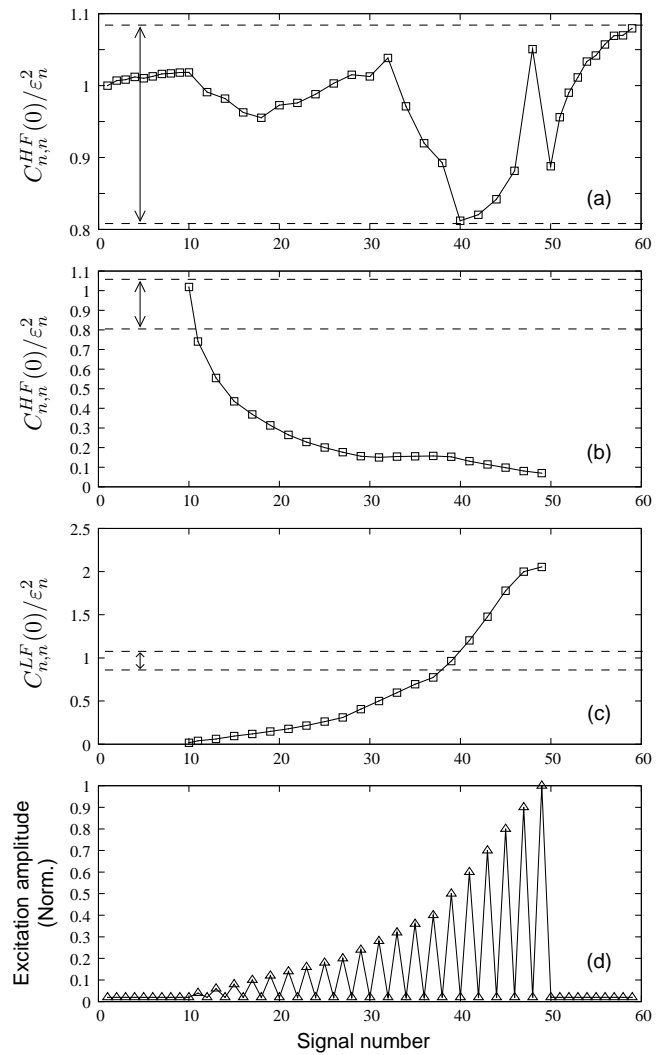}
\caption{(a) Energy of the high-pass filtered signals $C_{n,n}^{HF}(0)$ normalized by a quantity proportional to the excitation energy $\varepsilon_n^2$ for the low excitation amplitudes of the protocol (d). (b) Energy of the high-pass filtered signals for the higher excitation amplitudes of the protocol. (c) Energy of the low-pass filtered signals for the higher excitation amplitudes of the protocol. (d) Protocol for the successive excitation amplitudes of the experiment.}
\label{fig10}
\end{center}
\end{figure}

In order to analyze these amplitude dependent effects on the signal
energy attenuation, it is adequate now to focus on the energetical
properties of the signals. In Fig.~\ref{fig10}(d), the experimental
amplitude protocol is recalled. The total energies of the high-pass
filtered signals $C_{n,n}^{HF}(0)$ are normalized by a quantity
proportional to the excitation energy $\varepsilon_n^2$ both for the low excitation amplitudes of the protocol (Fig.~\ref{fig10}(a)) and for the higher excitation amplitudes (Fig.~\ref{fig10}(b)). The parts comprised between the dashed-lines compare the same range of normalized energy, from normalized values 0.8 to 1.1 roughly, that are the extreme values taken by the signals excited with the weak probe amplitude. The lowest value observed in Fig.~\ref{fig10}(a) corresponds to an excitation amplitude range where the parameter $R_{n,1}$ begins to decrease from 1 to a lower value. It is important to note that for the normalized energies associated with the higher excitation amplitudes, in Fig.~\ref{fig10}(b), all the values are below the region delimited by the dashed lines, which means that energy absorption is always stronger for $\varepsilon_a> \varepsilon_1$ even when the acoustic response of the medium has been modified by a strong wave action (even signals $s_{40}$ to $s_{50}$ for instance). The general behavior of the normalized energy for increasing excitation amplitudes is decreasing and monotonous: the higher the excitation amplitude is, the higher the energy absorption is. Its associated energy is consequently deviating from the $\varepsilon_n^2$ law. The nonlinear energy of the LF demodulated part is increasing monotonously with the excitation amplitude. The LF energy exceeds the HF energy of the signal from the excitation amplitude $\varepsilon_{41}$, as seen in Fig.~\ref{fig10}(c) when the LF energy is above the region delimited by the dashed lines.

\begin{figure}
\begin{center}
\includegraphics[width=8.5cm]{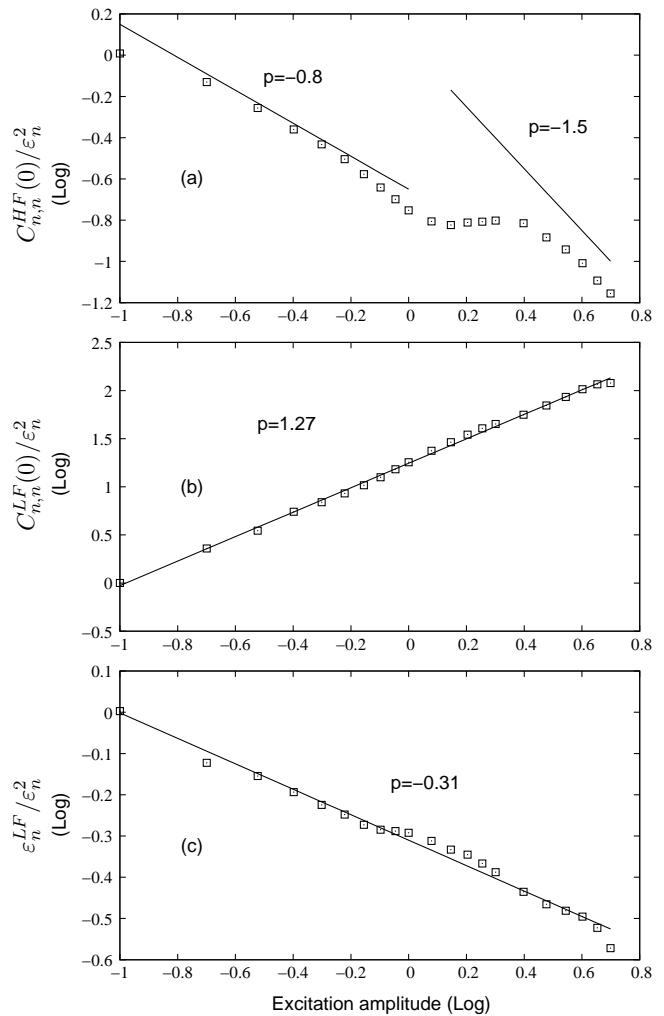}
\caption{(a) Energy of the high-pass filtered signals $C_{n,n}^{HF}(0)$ normalized by a quantity proportional to the excitation energy $\varepsilon_n^2$ as a function of the excitation amplitude in log-log scale. (b) Energy of the low-pass filtered signals $C_{n,n}^{LF}(0)$ normalized by a quantity proportional to the excitation energy $\varepsilon_n^2$ as a function of the excitation amplitude in log-log scale (c) Peak-to-peak amplitude of the self-demodulated signal normalized by $\varepsilon_n^2$ in log-log scale.}
\label{fig11}
\end{center}
\end{figure}

In order to analyze more precisely the nonlinear self-action process
for the HF components and the LF energy nonlinear increase process,
their normalized energies are plotted as a function of the excitation
amplitude in log-log scale, and compared to power laws in
Fig.~\ref{fig11}. In Fig.~\ref{fig11}(a), the normalized HF energy is
plotted as a function of the excitation amplitude. This is the same
quantity as that of Fig.~\ref{fig10}(b). For the lowest excitation
amplitudes, the normalized $C_{n,n}^{HF}(0)/\varepsilon_n^2$ is
decreasing close to a power law $\varepsilon_a^{-0.8}$, which means
that the energy $C_{n,n}^{HF}(0)$ is increasing with a
$\varepsilon_a^{1.2}$ dependence. In a perfectly linear system, it
would be a $\varepsilon_a^2$ dependence. Note that the overall
behavior is close to a $\varepsilon_a^{1.25}$ dependence, which is a
dramatic deviation from the linear law $\varepsilon_a^2$ and also a
very strong self-action compared to usually observed saturation
processes in rocks for instance \cite{whi83}. For higher excitation
amplitudes, a small plateau is observed, denoting a local comportment
law close to the linear one $\varepsilon_a^2$, even if the process
leading to this locally linear behavior may be due to combined effects
of different nonlinear processes: self-transparency, slow-dynamics, conditioning...
For the highest excitation amplitudes, the dependence on amplitude is
the strongest, approaching a $\varepsilon_a^{0.5}$ law for
$C_{n,n}^{HF}(0)$. This means that if the excitation energy is
multiplied by a factor $4$ for instance, the received energy is
multiplied by a factor $\sqrt{2}$. This would be an even stronger
effect for the late part of the signal considering the fact that most
of the nonlinearly attenuated energy is the one arriving at the latest
time in the signal, according to Fig.~\ref{fig7} and
Fig.~\ref{fig8}. Surprisingly, the largest amplitudes of the signal
are not strongly nonlinearly attenuated but the late contributions are. This peculiar observation is discussed in section \ref{sec4}.

In Fig.~\ref{fig11}(b), the normalized LF part energy
$C_{n,n}^{LF}(0)/\varepsilon_n^2$ is plotted in log-log scale as a
function of the excitation amplitude. A quite clear power law
$\varepsilon_a^{1.27}$ is found for this NL process of frequency-down
conversion (self-demodulation), which corresponds to a LF energy
proportional to $\varepsilon_a^{3.27}$. The classical NL behavior
observed in homogeneous media with a quadratic nonlinearity is
$\varepsilon_a^4$ for the LF energy \cite{nov87}. The observed
dependence corresponds to a $\varepsilon_a^{1.63}$ power law for the
field itself, to be compared to the $\varepsilon_a^2$ law classically
observed. This dependence is also closely observed with another
processing method applied to the LF part: the normalized peak-to-peak
amplitude of the LF contribution $\varepsilon_n^{LF}/\varepsilon_n^2$
is plotted in Fig.~\ref{fig11}(c) in log-log scale. A power law
behavior $\varepsilon_a^{-0.31}$ can be observed, that is a law
$\varepsilon_a^{1.69}$ for the dynamics of $\varepsilon_n^{LF}$. If
compared to the behavior obtained with $C_{n,n}^{LF}(0)\propto
\varepsilon_a^{3.27}$, taking into account that $C_{n,n}^{LF}(0)
\propto (\varepsilon_n^{LH})^2$, this dynamics corresponds to a
$\varepsilon_a^{1.69 \times 2} = \varepsilon_a^{3.38}$ law. This is an
intermediate dependence between the classical one $\varepsilon_a^4$
and the Hertzian clapping limit $\varepsilon_a^3$. The latter dependence law has been observed earlier in different experimental conditions in \cite{tou04a}. Note that for symmetry reasons, the Hertz-Mindlin hysteretic nonlinearity is not able to predict the effect of self-demodulation observed here at the leading order of the nonlinear approximation, and may not contribute \cite{zar71,gus98}. 

\section{Summary of the results and discussions}\label{sec4}

Different experimental observations have been presented for the first time and the main results can be summarized as the following:
\begin{itemize}
\item observation of the coexistence of a nonlinear self-demodulated LF wave together with a HF coda-type wave (Fig.~\ref{fig2}),
\item a drastic increase of the HF coda energy with static stress, compared to the dependence of the LF wave energy (Fig.~\ref{fig5}),
\item  a strong shape dependence of the coda on the acoustic excitation amplitude (Fig.~\ref{fig6} and Fig.~\ref{fig7}), and especially a strong nonlinear attenuation of the coda wave at the latest arrival times,
\item almost no variation in the acoustic properties of the medium
  after moderate acoustic excitations up to an excitation acoustic strain of
  $\varepsilon_a \sim 0.3 \times 10^{-5}$, corresponding to signal number 31
  (this represents a ratio of dynamic over static deformation of
  $\varepsilon_a/\varepsilon_0 \simeq 5\times 10^{-3}$),
\item dynamic modifications of the acoustic properties (nonlinear
  self-action, fast dynamics) even at low excitation levels
  $\varepsilon_a \sim 2 \times 10^{-7}$ corresponding to signal number 11 (this represents a ratio of dynamic over static deformation of $\varepsilon_a/\varepsilon_0 \simeq 3\times 10^{-4}$),
\item  modification of the acoustic properties after large excitation levels (slow dynamics and conditioning effects), followed by a slow recovery.
\end{itemize}

Having in mind the ray acoustic approximation in this experiment of multiple scattering of elastic waves, the first part of the coda signals can be associated with short propagation paths while the last part of the coda signals may be associated with longer propagation paths. Note that the lengths of the long propagation paths can exceed the container size by a factor of 10, which means that reflexions on the container sides may occur. In principle, nonlinear attenuation should be more visible on the long propagation paths than on the short ones because the distance for the accumulation of this nonlinear process is larger \cite{zar71,nov87}. This explains the stronger effect of nonlinear attenuation on the last part of coda signals, that experienced longer propagation paths. The physical nature of this nonlinear attenuation process remains undetermined. Nonlinear scattering of the elastic waves due to opening or closing of the weakest contacts and therefore dynamic switching of propagation paths under the wave action could explain the nonlinear attenuation. This is in agreement with the drastic dependence of the HF coda wave energy transmission on the applied static stress. Nonlinear absorption of the elastic waves at the contacts by stick-slip \cite{min53} or thermo-elasticity \cite{zai03} processes could also contribute. Recent numerical simulations of the acoustic propagation through unconsolidated granular packings \cite{lud05,mou08} could provide an interesting insight in the processes involved in the experimentally observed nonlinear attenuation.

For both nonlinear scattering and nonlinear absorption, the weakest
contacts are supposed to be responsible for the main contribution,
especially for such small acoustic excitation strains as $\varepsilon_a
\simeq 10^{-7}$. Coda signals are shown to be particularly sensitive
to these weak contacts, certainly due to the fact that energy is well
distributed over the entire medium, including weak contacts, after few
milliseconds (coda signals have durations larger than $2\ ms$,
which corresponds to propagation paths much larger than the emitter -
receiver distance). Moreover, the wavelength being of the order of the
bead diameter ($\lambda \sim d$), small features as small as the beads
and their contacts are resolved by the acoustic wave. This is not the case for long wavelength propagation ($\lambda \gg d$), which is known to be less sensitive to the weak contacts in the medium and more influenced by the average properties of the medium \cite{zai05,tou04a}.

The modification of the acoustic properties after the application of a
strong excitation, is commonly related to the effect of conditioning
and slow dynamics \cite{hol81,joh05,joh05b,zai03}. This addresses the
problem of the acoustic wave influence on the granular medium state
and memory. Under a given threshold, the medium is apparently not
modified (under excitation amplitude number 31, $\varepsilon_a \simeq
0.3\ 10^{-5}$ which corresponds to a ratio of dynamic over static deformation of
  $\varepsilon_a/\varepsilon_0 \simeq 5\times 10^{-3}$), while over some excitation amplitude, the medium exhibits memory of the past acoustic excitation. Coda signal analysis is shown to be a powerful tool to monitor slight modifications in the acoustic response of an unconsolidated granular structure as a function of time, the resolution (including some time averaging) being of the order of the second.

\section{Conclusions}

In this work, experimental results on the nonlinear transport of short
wavelength acoustic waves (of the order of the bead diameter) are
reported. It is shown that the amplitude dependence of the codas is
strong, and contains some information on the strength of the
propagation paths in the medium. An amplitude dependent attenuation is
visible on the characteristic time $T_e$, which diminishes with the
excitation amplitude: most of the nonlinear self-action effects take
place at the latest arrival times of the coda signals. This shows that
the field which encountered the larger number of scattering events, or which does not follow the strongest force chains is more influenced by the effects of nonlinear dissipation (stick-slip, sliding, or thermo-elastic losses) and nonlinear scattering (closing or opening of the weak contacts for instance).
 
Among the possible applications of such experimental results is the non destructive testing (NDT) of granular structures with nonlinear acoustic methods, in particular with multiple scattered waves (coda signals), in contrast to the widely used and recent nonlinear methods employing coherent wave interactions \cite{zhe99,jac03}. For this purpose, it would be of interest to make use in the future of the coda wave interferometry technique \cite{sni02,lob03,sni06}. 

It has been shown that the acoustic properties of the medium can be modified temporary by a strong acoustic excitation. In the context of NDT, this work show the range of acoustic amplitudes to use (or the time necessary to recover the initial state) in order to stay in a regime where the acoustic waves do not modify the elastic response of the medium.

\begin{acknowledgments}
This work is supported by ANR project ``grANuLar'' NT05-3\underline{\ }41989. The authors would like to thank E. Brasseur, E. Egon and P. Collas for their technical help on the experimental setup.
\end{acknowledgments}


\end{document}